\documentclass[prb,twocolumn,amsmath,amssymb,superscriptaddress,floatfix,nofootinbib]{revtex4-2}

\usepackage{graphicx}
\usepackage{dcolumn}
\usepackage{bm}
\usepackage{ textcomp }
\usepackage{color}
\usepackage{amsmath}
\usepackage{amssymb}
\usepackage{xcolor}
\usepackage[colorlinks = true,
            linkcolor = blue,
            urlcolor  = red,
            citecolor = blue,
            anchorcolor = blue]{hyperref}
\usepackage{easyReview}
\usepackage{mathdots}
\usepackage{float}
\usepackage{lineno}
\usepackage{todonotes}
\usepackage{url}
\usepackage{array}
\usepackage{todonotes}
\usepackage{xcolor,colortbl}

\usepackage{lipsum, babel}

\definecolor{LightBlue}{rgb}{0.8,0.8,0.8}
\newcommand{\pdag}{{\phantom\dagger}}

\usepackage[normalem]{ulem}

\begin{document}
\title{Superconductivity in the bilayer Hubbard model: Are two Fermi surfaces better than one?}
\date{\today}

\author{Seher Karakuzu}
\affiliation{Computational Sciences and Engineering Division, Oak Ridge National Laboratory, Oak Ridge, Tennessee 37831-6164, USA}

\author{Steven Johnston}
\affiliation{Department of Physics and Astronomy, The University of Tennessee, Knoxville, Tennessee 37966, USA}

\author{Thomas A. Maier}
\affiliation{Computational Sciences and Engineering Division, Oak Ridge National Laboratory, Oak Ridge, Tennessee 37831-6164, USA}

\begin{abstract}
Fully occupied or unoccupied bands in a solid are often considered inert and irrelevant to a material's low-energy properties. But the discovery of enhanced superconductivity in heavily electron-doped FeSe-derived superconductors poses questions about the possible role of incipient bands (those laying close to but not crossing the Fermi level) in pairing. To answer this question, researchers have studied pairing correlations in the bilayer Hubbard model, which has an incipient band for large interlayer hopping $t_\perp$, using many-body perturbation theory and variational methods. They have generally found that superconductivity is enhanced as one of the bands approaches the Liftshiz transition and even when it becomes incipient. Here, we address this question using the nonperturbative quantum Monte Carlo (QMC) dynamical cluster approximation (DCA) to study the bilayer Hubbard model's pairing correlations. We find that the model has robust $s_\pm$ pairing correlations in the large $t_\perp$ limit, which can become stronger as one band is made incipient. While this behavior is linked to changes in the effective interaction, we further find that it is  counteracted by a suppression of the intrinsic pair-field susceptibility and does not translate to an increased $T_c$. Our results demonstrate that the highest achievable transition temperatures in the bilayer Hubbard model occur when the system has two bands crossing the Fermi level.  
\end{abstract}
\maketitle

\section{Introduction}
To harness the full potential of superconductors for technological applications, we must develop methods to engineer and optimize properties like their transition temperature $T_c$ and critical current density $J_c$. The wide range of $T_c$ values achieved in the FeSe-derived family of high-temperature (high-$T_c$) superconductors \cite{KreiselSymmetry2020, Hsu14262, WangCPL2012, NOJI20148, BurrardLucasNatMat2013, DingPRL2016, Liu2012, He2013, DingNatCommun2013, Ge2015, Shi2017, GuoPRB2010, PhysRevB.80.064506, Medvedev2009, MiyataNatMat2015, SongPRL2016, Krzton_Maziopa_2012, Sun2015} has attracted considerable attention in this context, as understanding the mechanisms behind this tunability could help uncover general principles for engineering superconductivity.   

Bulk FeSe is an unconventional superconductor with $T_c\approx 8$ K at ambient pressure~\cite{Hsu14262}. Its electronic structure in bulk form resembles other Fe-based superconductors with a set of hole-like bands crossing the Fermi energy $E_\mathrm{F}$ at the $\Gamma$ point [${\bf k} = (0,0)$] and another set of electron-like bands crossing at the $M$-point [${\bf k} = (\pi/a,\pi/a)$, two-Fe unit cell notation]. The system can be electron-doped by intercalating alkali atoms \cite{GuoPRB2010, DingNatCommun2013} or more complex molecules \cite{NOJI20148, BurrardLucasNatMat2013, Krzton_Maziopa_2012, Sun2015} into the van der Waals gap between the layers. This process can induce a Lifshitz transition whereby the hole-like bands sink below $E_\mathrm{F}$, leaving only the electron-like Fermi surface pockets at the zone corner. At the same time, $T_c$ increases to $\approx 45$ K. Growing monolayers of FeSe on oxide substrates like SrTiO$_3$, BaTiO$_3$, or TiO$_2$ also electron dopes the system~\cite{Liu2012} but one can achieve $T_c \sim 55-75$~K in this case, even for electron concentrations comparable to those realized in FeSe intercalates~\cite{WangCPL2012, Liu2012, He2013, Ge2015}. 

On the one hand, the observation of enhanced superconductivity in FeSe monolayers on oxide substrates indicates that the substrate contributes to raising $T_c$ in FeSe monolayers \cite{LeeNature2014, Lee2015, Rademaker2016, SongNatureCommun2019, RademakerEnhanced2021}. This possibility has renewed efforts towards engineering unconventional superconductivity in artificial heterostructures, and interfaces \cite{LeeNature2014, CohHeterostructure, PengNatureCommun2014}. On the other hand, the increased $T_c$ found in the FeSe intercalates provides compelling evidence that electron doping and the associated changes in the electronic band structure also play a critical role in establishing high-$T_c$ values in these materials. While both aspects are interesting in their own right, here we focus our attention on the latter. 

An early and influential picture for superconductivity in the iron-based superconductors \cite{Mazin2008} was built around the notion that spin fluctuations meditate electron pairing, where strong nesting between electron and hole pockets leads to an $s^\pm$ gap symmetry~\cite{Kuroki2008, Mazin2008, Hirschfeld_2011}. At face value, the existence of a system with an \emph{increased} $T_c$ but with its hole-like bands laying below $E_\mathrm{F}$ (i.e., an incipient band\footnote{The term ``incipient band'' seems to have multiple usages in the literature. Here, we use the term to refer to a band that is full (empty) but whose maximum (minimum) is close to the Fermi energy.}) challenges this picture. These observations have led to an effort to understand how bands close to \cite{Hirschfeld_2011, PhysRevB.93.155159, PhysRevB.91.161108, BangNJP2014, ChenPRB2015, Mishra2016, LinscheidPRL2016, KurokiFlex2020, KurokiVMC2020, Maier2019, RademakerEnhanced2021} (or even far from \cite{PhysRevB.97.060501}) the Fermi energy might affect superconductivity. 

While researchers have studied several models for incipient band systems to date, the focus of this work is on the bilayer Hubbard model. It is a simple two-band model whose Fermi surface can be tuned continuously between having two disconnected sheets to a system with a single sheet and an incipient band, in analogy to the FeSe family~\cite{Maier2011}. Moreover, the model retains only the on-site Hubbard interactions, making it easier to solve using nonperturbative methods like quantum Monte Carlo (QMC)~\cite{Maier2011, Maier2019, PelliciariRIXS2020, KurokiVMC2020},  where interorbital interactions generally produce severe Fermion sign problems.

\begin{figure*}[t]
    \centering
    \includegraphics[width=\textwidth]{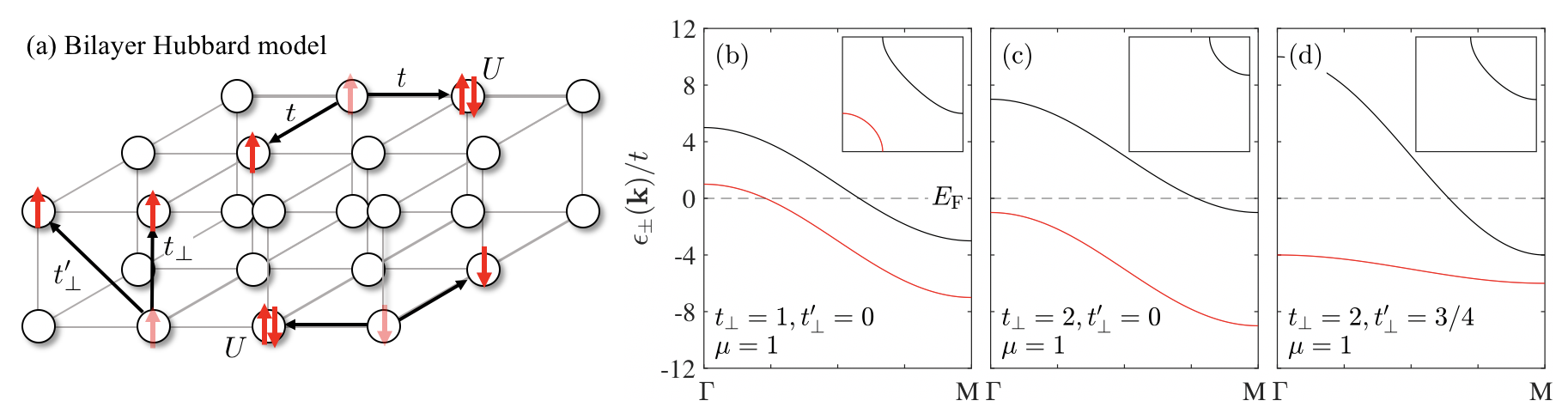}
    \caption{(a) A sketch of the bilayer Hubbard model and the hopping processes considered in this work. The model includes nearest-neighbor intralayer hopping $t$ as well as nearest- and next-nearest-neighbor interlayer hopping $t^{\phantom\dagger}_\perp$ and $t^\prime_\perp$, respectively. The model also includes an onsite Hubbard repulsion $U$ in each layer. Panels (b)-(d) show the noninteracting band structure $\epsilon_\pm({\bf k})$ along the Brillouin zone diagonal from $\Gamma$ [${\bf k } = (0,0)$] to $M$ [${\bf k} = (\pi/a,\pi/a)]$ for various model parameters, as indicated in each panel. The dashed line indicates the position of the Fermi energy $E_\mathrm{F}$. The insets in (b)-(d) show the corresponding Fermi 
    surface(s) for each parameter set.}
    \label{fig:bilayer_sketch}
\end{figure*}

Remarkably, recent weak coupling calculations for this model within a fluctuation exchange (FLEX) approximation have found that superconductivity is strongly enhanced when one of the bands is close to but still below the Fermi energy~\cite{KurokiFlex2020}. The authors attributed this enhancement to a suppression of the low-energy spin fluctuations due to the incipient nature of the hole band and the associated transfer of spectral weight to an energy range optimal for pairing. Later, the same group investigated this question in the strong-coupling regime using variational Monte Carlo (VMC) calculations and found that the zero-temperature pairing correlations are more robust in a regime where one of the bands is incipient~\cite{KurokiVMC2020}, consistent with their earlier work. 

The above results suggest that pairing and the superconducting $T_c$ can be optimized in the bilayer Hubbard model by shifting one band close to or below the Fermi energy. However, as a many-body perturbation theory method, FLEX approximates the electronic interactions and can have difficulties describing a model's strong coupling regime in a controlled way. On the other hand, VMC can treat strong interactions, but the quality of the variational wave functions can limit its accuracy. Moreover, VMC cannot access excited state properties or treat the system in the thermodynamic limit. Whether or not the bilayer Hubbard model has an enhanced superconducting $T_c$ close to or across the Lifshitz transition, therefore, remains an open question. 

We address this issue by using dynamical cluster approximation (DCA) calculations~\cite{Jarrell2001, Maier_Review} with a QMC solver to study how superconductivity evolves in the bilayer model. We find that the model has robust $s^\pm$ pairing correlations in the large $t_\perp$ limit, both when two bands cross $E_\mathrm{F}$ or when one is incipient. Moreover, the strength of the pairing correlations can be enhanced as the bandwidth of the nearly-incipient band is narrowed, provided the band still crosses $E_\mathrm{F}$. However, we find that the value of $T_c$ ultimately realized in the system is consistently and significantly suppressed once the system undergoes a Lifshitz transition and one band becomes fully submerged. This behavior is linked to increases in the effective interaction that are counteracted by a suppression of the intrinsic pair-field susceptibility. We also demonstrate that the largest $T_c$ in the bilayer model with large interlayer hopping is achieved when both bands cross $E_\mathrm{F}$. While these results can provide some intuition into the mechanisms at play in materials like the FeSe-derived high-$T_c$ superconductors, these comparisons must be made with several caveats that we will discuss in Sec.~\ref{sec:discussion}. We note, however, that the bilayer Hubbard model has recently been realized in cold atom experiments~\cite{GallNature2021}, which can provide crucial experimental verification of our results. 

\section{Model and Methods}\label{Sec:Methods}
\subsection{The Bilayer Hubbard Model}
We consider the bilayer Hubbard model (Fig.~\ref{fig:bilayer_sketch}a) defined on a two-dimensional square lattice with $L^2$ unit cells (or $N = L\times L \times 2$ orbitals), where $L$ is the linear size of the system. The Hamiltonian is given by
\begin{eqnarray}
  {\cal H} = {\cal H}_{\cal K} + {\cal H}_{\cal U},
  \label{eq:Hmodel}
\end{eqnarray}
where
\begin{eqnarray}
  {\cal H}_{\cal K}&=&- \mu \sum_{i,\alpha,\sigma} n_{i,\alpha,\sigma} + \sum_{\substack{i,j\nonumber\\ \alpha, \sigma}} 
  t^{\phantom\dagger}_{i,j}\left(c^\dag_{i,\alpha,\sigma} c^\pdag_{j,\alpha,\sigma}  + {\rm H. c.} \right) \\
  &+& \sum_{i,j,\sigma} t^{\perp}_{i,j} \left(c^\dag_{i,1,\sigma} c^\pdag_{i,2,\sigma}  + {\rm H. c.} \right) 
  \label{eq:Hk}
\end{eqnarray}
and
\begin{equation}
  {\cal H}_{\cal U} =  U \sum_{i,\alpha} n_{i,\alpha,\uparrow} n_{i,\alpha,\downarrow}.
  \label{eq:HU}
\end{equation}  
Here, $c^\dag_{i,\alpha,\sigma}$ ($c^\pdag_{i,\alpha,\sigma}$) creates (annihilates) a spin $\sigma$ ($=\uparrow,\downarrow$) electron in the unit cell $i$ and layer $\alpha$ ($=1,2$), $t_{i,j}$ and $t^\perp_{i,j}$ are the intra- and interlayer hopping integrals, respectively,  $n^{\phantom\dagger}_{i,\alpha,\sigma} = c^\dag_{i,\alpha,\sigma} c^{\phantom\dagger}_{i,\alpha,\sigma}$ is the number operator, and $U$ is the Hubbard interaction, which acts only between electrons in the same layer. The average particle number per orbital $n = \frac{1}{2}\sum_{i,\alpha,\sigma}\langle n_{i,\alpha,\sigma}\rangle$ is controlled by the chemical potential term $\mu$. Throughout, we set the intralayer hopping  $t_{i,j} = t$ for in-plane nearest neighbors and $t_{i,j} = 0$ otherwise, and set $t = 1$ as our unit of energy. For the interlayer hopping, we set $t^\perp_{i,j} = t_\perp$ for $i = j$ and $t^{\perp}_{i,j} = t_\perp^{\prime}$ for next-nearest inter-plane neighbors $\langle i,~j\rangle$, and zero otherwise, as sketched in Fig.~\ref{fig:bilayer_sketch}a. The majority of our results for $t_\perp^\prime = 0$ are obtained on $N = 4\times 4\times 2$ clusters, unless otherwise stated. We have found that $t^\prime_\perp \ne 0$ produces a more significant Fermion sign problem and so most of results for this case were obtained on $2\times 2\times 2$ clusters, unless otherwise stated. 

\subsection{Method Details}
We study the bilayer Hamiltonian in Eq.~(\ref{eq:Hmodel}) using the DCA method~\cite{Jarrell2001, Maier_Review} with a 
continuous-time auxiliary field QMC (CT-AUX) cluster solver~\cite{Gull_2008, Gull2011}.   
DCA is an embedded cluster method, where the infinite system is mapped onto a finite size cluster and then solved self consistently. The algorithm treats the correlations in the cluster exactly while approximating correlations on longer length scales using a dynamic mean-field. The Fermion sign problem for the QMC solver is less severe within the DCA approach. This aspect allows us to simulate larger clusters and lower temperatures than other finite cluster algorithms~\cite{Jarrell2001}. We used the DCA++ implementation of the DCA algorithm in this work, 
as detailed in Ref.~\citenum{HAHNER2019}. 

\subsection{Observables}
To investigate the strength of the pairing correlations and the symmetry of the dominant superconducting correlations, we have solved the Bethe-Salpeter equation (BSE) in the particle-particle channel \cite{Maier2006}
\begin{equation}
    \lambda_\nu(T) \phi_\nu(k) = - \frac{T}{N} \sum_{k^\prime} \Gamma(k,k^\prime) G(k^\prime) G(-k^\prime) \phi_\nu(k^\prime).
    \label{eq:BSEeq}
\end{equation}
Here, $k\equiv(k_x,k_y,k_z,w_n)$ with $w_n=(2n+1)\pi T$ a Fermion Matsubara frequency, $G(k)$ is the fully dressed single particle propagator, and $\Gamma(k,k^\prime)$ is the particle-particle irreducible vertex function. The system undergoes a superconducting transition at $T=T_c$ when the leading eigenvalue $\lambda_0(T_c) = 1$ and the symmetry of the superconducting state is determined from the momentum and frequency structure of the corresponding eigenfunction $\phi_0(k)$.

To help understand the evolution of the pairing correlations with model parameters, we can define the intrinsic (non-interacting dressed) pair-field susceptibility as in Ref. \citenum{PhysRevResearch.2.033132} 
by projecting onto the leading eigenfunction $\phi_0(k)$ 
\begin{equation}\label{eq:P0}
    P_0(T) = \frac{T}{N} \sum_k \phi_0^2(k) G(k) G(-k).
\end{equation}
After doing so, an effective interaction strength can be defined as 
\begin{equation}\label{eq:V}
    V_0(T) = \lambda_0(T)/P_0(T).
\end{equation}

We also investigate the spin correlations in the system by calculating the dynamical spin susceptibility 
\begin{equation}
    \chi^{\alpha,\beta}_{s}(\mathbf{r}_i-\mathbf{r}_j,\tau) = \langle {\cal T}_\tau S_{i,\alpha}^{z}(\tau)S_{j,\beta}^{z}(0)\rangle,
\end{equation}
where ${\cal T}_\tau$ is the time ordering operator, and  $S_{i,\alpha}^{z}(\tau)= \frac{1}{2}\left[n_{i,\alpha,\uparrow}(\tau)-n_{i,\alpha,\downarrow}(\tau)\right]$ is the $z$-component of the spin on orbital (layer) $\alpha$ at site $\mathbf{r}_i$ at time $\tau$. A Fourier transform of the real space spin susceptibility will result in the momentum dependent zero frequency susceptibility 
\begin{equation}\label{eq:spinsus}
    \chi_{s}(\mathbf{Q}) = \sum_{\mathbf{r},\alpha,\beta} \int\limits_0^\beta d\tau\, \chi_{\alpha,\beta}^{s}(\mathbf{r},\tau) e^{i\mathbf{Q_{\parallel}}\cdot \mathbf{r}} e^{iQ_{\perp}(r_{z,\alpha}-r_{z,\beta})} ,
\end{equation}
where $\mathbf{Q} = (\mathbf{Q}_{\parallel}, Q_{\perp}) = (Q_x,Q_y, Q_z)$ is the scattering wavevector and $r_{z,\alpha} = 0,~1$ for orbitals in layers $\alpha = 0,~1$.

\section{Results}
\subsection{The noninteracting band structure}

We first present the effect of Hamiltonian parameters $t^{\phantom\prime}_\perp$ and $t_\perp^\prime$ on the electronic structure of the model. In the noninteracting limit ($U = 0$), Eq.~(\ref{eq:Hmodel}) can be diagonalized exactly in momentum space, where the bands are given by even ($+$) and odd ($-$) combinations of the orbitals in each layer \cite{KurokiFlex2020}
\begin{equation}\label{eq:epsilon0}
    \epsilon_\pm({\bf k})= 2(t\pm t_\perp^\prime)\left[\cos(k_xa)+\cos(k_ya)\right] \pm  t_\perp - \mu. 
\end{equation}
Here, $a=1$ is the in-plane lattice constant and our unit of length. The $(+)$ and $(-)$ combinations can also be viewed as Bloch states of a one-band model with $k_z = 0$ and $\pi$, respectively. We will use this notation for the remainder of this work. 

The noninteracting band structure for different choices of interlayer hopping $t^{\phantom\prime}_\perp$ and $t_\perp^\prime$ and $\mu = t$ are plotted in Fig.~\ref{fig:bilayer_sketch}(b)-(d). Increasing the value of $t_\perp$ controls the energy separation between the bands, as shown in Figs.~\ref{fig:bilayer_sketch}(b) and \ref{fig:bilayer_sketch}(c). Increasing $t_\perp^\prime > 0$ increases (decreases) the bandwidth of the $k_z = 0$ ($\pi$) bands, as shown in Fig.~\ref{fig:bilayer_sketch}(d). Importantly, the lower-energy band can be driven through a Lifzhits transition for sufficiently large values of $t^{\phantom\prime}_\perp$ and/or $t_\perp^\prime$ and suitable values of the chemical potential $\mu$. Across this transition, the system goes from having two disconnected Fermi surfaces centered at $\Gamma$ and $M$ to having only a single Fermi surface at $M$ and an incipient band at $\Gamma$, as sketched in the insets of \ref{fig:bilayer_sketch}(b)-(d). 
We can, therefore, adjust the interlayer hopping parameters to tune the system between these cases. 
However, the Hubbard interaction will further renormalize the bands, which can alter their relative positions. For example, the Hartree contribution to the electron self-energy $\Sigma({\bf k},\mathrm{i}\omega_n)$ will further separate the bands \cite{KurokiFlex2020, RademakerEnhanced2021} while other self-energy terms renormalize the electronic structure~\cite{KurokiFlex2020, RademakerEnhanced2021, Maier2019, PelliciariRIXS2020}. It can be the case that one of the renormalized bands is incipient, even when both noninteracting bands cross $E_\mathrm{F}$. It is, therefore, necessary to compute the dressed electronic structure as encoded in the single-particle spectral function $A({\bf k},\omega)$. Throughout, we obtained  $A({\bf k},\omega)$ by analytically continuing our results to the real frequency axis using the maximum entropy method~\cite{Gubernatis1991}. The reader can find additional details in Ref.~\citenum{Maier2019}. 

\subsection{Results without next-nearest-neighbor inter-layer hopping}

\begin{figure}[t]
\centering
\includegraphics[scale=0.34]{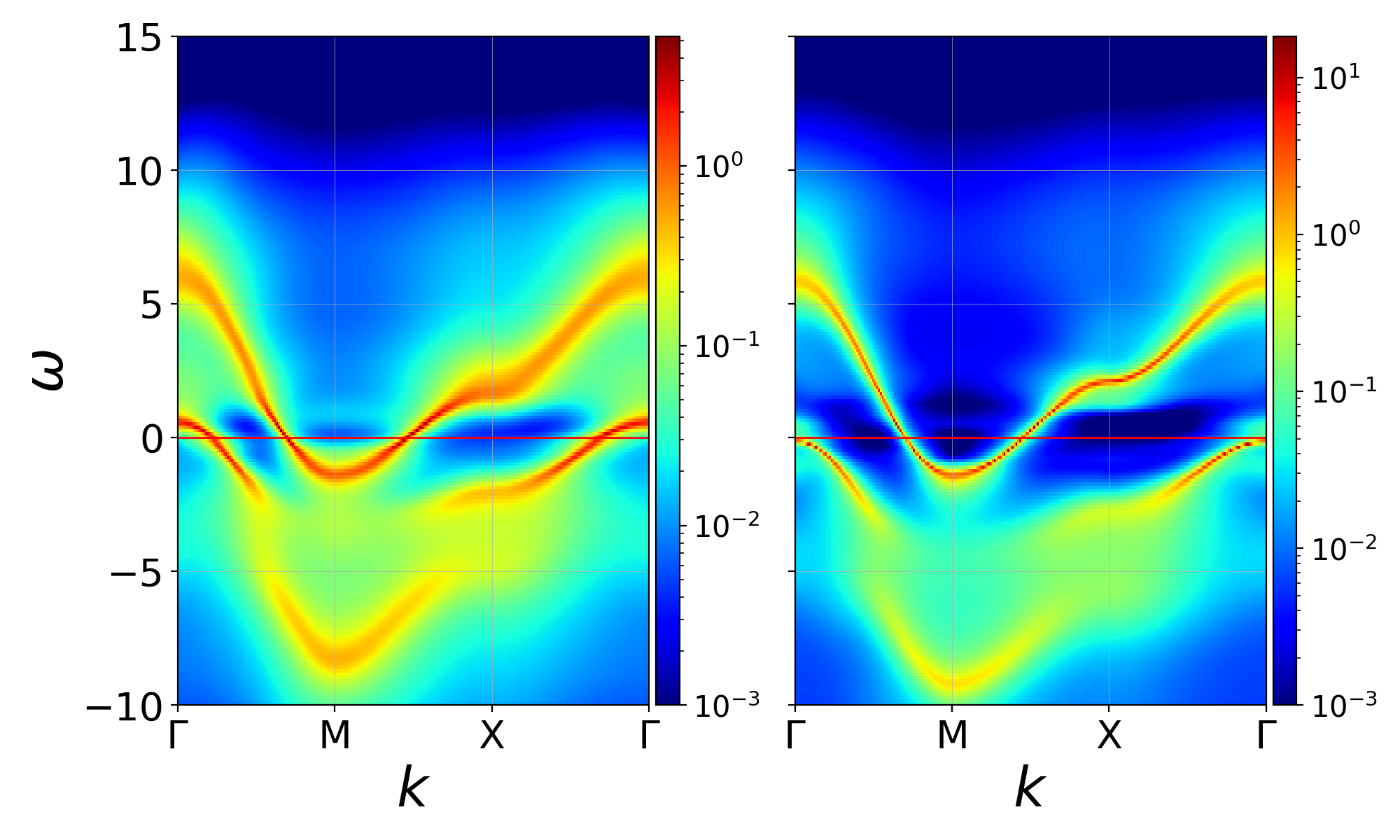}
\caption{The single-particle spectral function $A({\bf k},\omega)$ for the bilayer Hubbard model, computed at an inverse temperature $\beta = 5/t$ and on an $4\times4\times2$ cluster with $U = 6t$. The left panel shows results for $t_{\perp} = 2.3t$, $t_\perp^\prime = 0$, and $n =1.10$. The right panel panels hows results for $t_{\perp} = 2.8t$, $t_\perp^\prime = 0$, and $n =1.15$. The red line indicates the position of the Fermi energy. The high symmetry points are defined as $\Gamma = (0,0)$, $M = (\pi/a,\pi/a)$, and $X = (\pi/a,0)$.}
\label{fig:Akwbulkinc}
\end{figure}

\begin{figure}[t]
\centering
\includegraphics[scale=0.4]{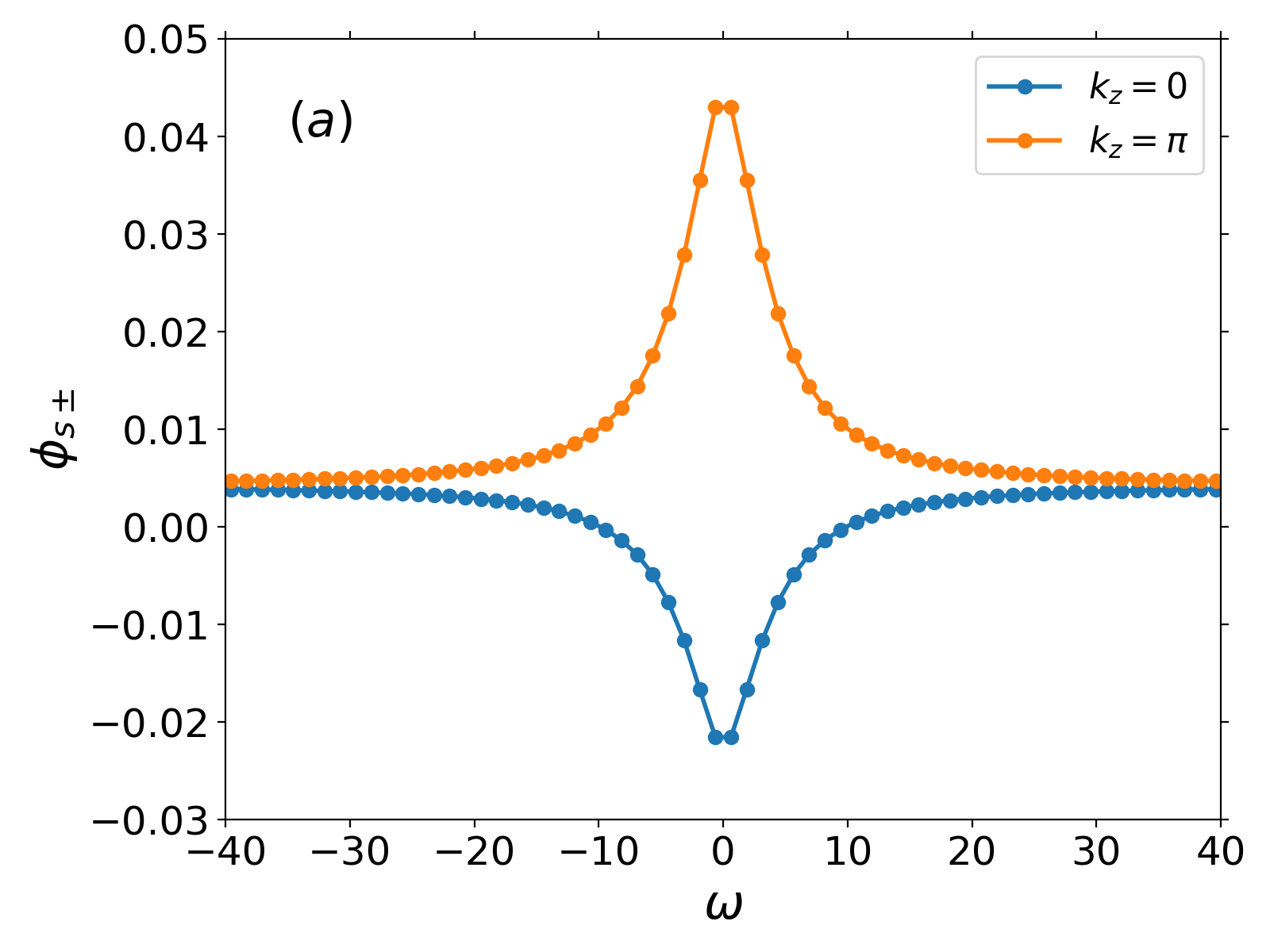}
\includegraphics[scale=0.4]{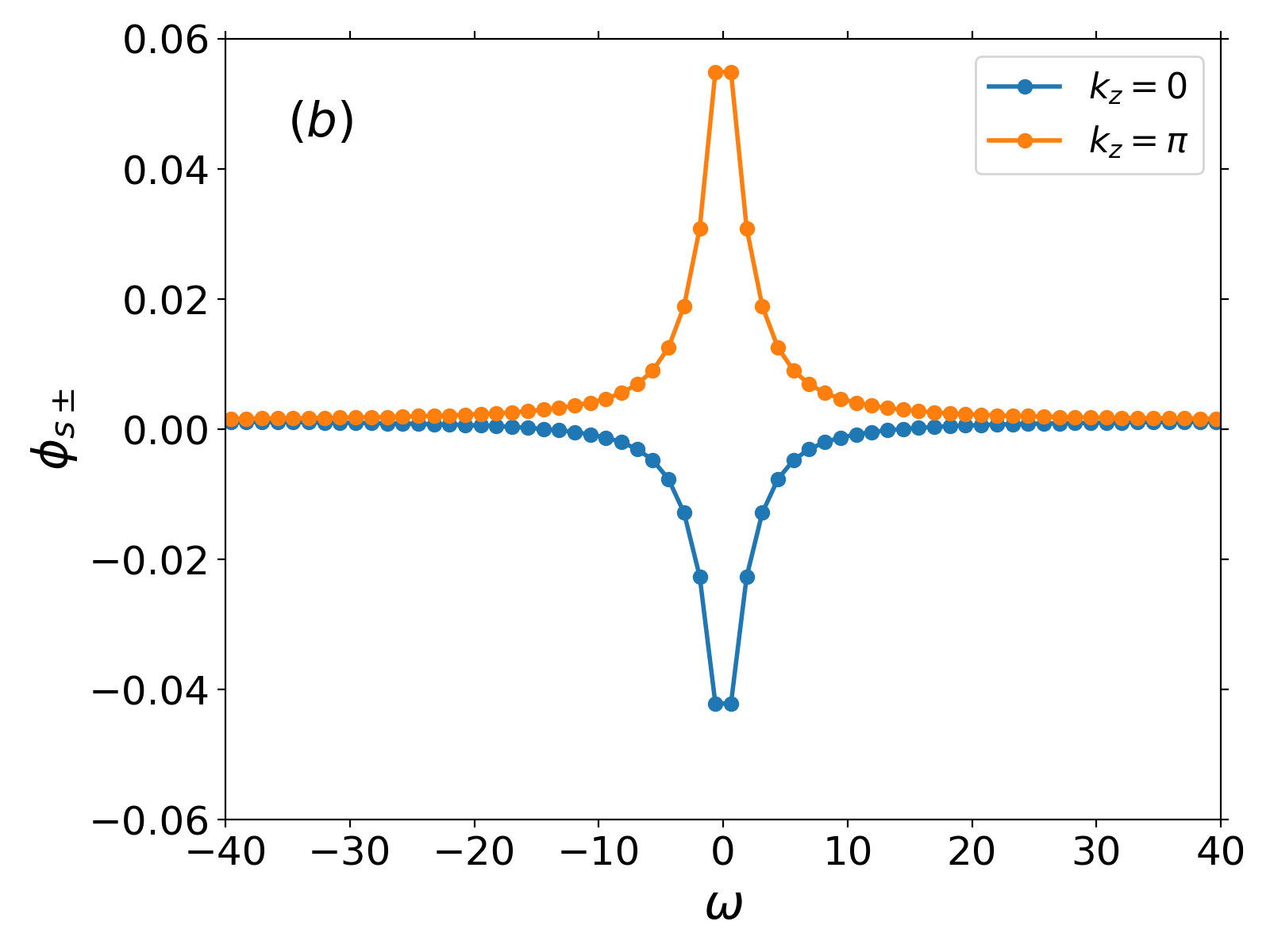}
\caption{The frequency dependence of the leading BSE eigenvectors, averaged over ${\bf k} = (k_x,k_y)$ in the first Brillouin zone, and projected onto the high- ($k_z = 0$) and low-energy ($\pi$) bands. Results are shown for (a) an incipient band case with $n=1.15$, $\beta=5t$, $t_{\perp}=2.8t$ , and $t_\perp^\prime = 0$, and (b) a two-band case $n=1.05$, $\beta=5t$, $t_{\perp}=2.3t$, and $t_\perp^\prime = 0$.}
\label{fig:kavEig}
\end{figure}

Figure \ref{fig:Akwbulkinc} plots the dressed single-particle spectral function $A({\bf k},\omega)$ along high-symmetry cuts of the Brillouin zone for a bilayer model with $U = 6t$. The left panel shows results for $t^{\phantom\prime}_\perp= 2.3t$, $t_\perp^{\prime}=0$, and an average density of $n=1.10$. In this case, the system has a band structure with two bands crossing the Fermi level forming an electron pocket at the $M$-point and a hole pocket at the $\Gamma$-point. The right panel shows results for $t^{\phantom\prime}_\perp= 2.8t$, $t_\perp^{\prime}=0$, and an average density of $n=1.15$. In this case, the system has just an electron pocket at the $M$ point while the hole band is pushed below $E_\mathrm{F}$. 

The results shown in Fig.~\ref{fig:Akwbulkinc} establish that we can indeed modify the Fermi surface topology by changing the density ($\mu$) and the perpendicular hopping $t_\perp$, even when $U \ne 0$. This result is consistent with prior nonperturbative results for this model~\cite{Maier2019, PelliciariRIXS2020}. For example, the band structures presented in Fig.~\ref{fig:Akwbulkinc}a and Fig.~\ref{fig:Akwbulkinc}b bear some resemblance to those observed in bulk and monolayer FeSe, respectively~\cite{PelliciariRIXS2020}. In this context, it is interesting to note that the spectral function in the incipient band case is significantly sharper than in the two-band case. (Note the change in the intensity scale between the left and right panels.) This difference indicates that the incipient band case exhibits weaker electronic correlations than the two band case, even though both have comparable values of $U/W$. This observation also agrees with more realistic LDA+DMFT treatments of FeSe~\cite{MandalPRL2017}, which also found that the spectral functions for the FeSe monolayer were sharper than those computed for the bulk.  

Next, we examine how the superconducting correlations evolve with the band structure topology by solving the BSE as formulated in Eq.~\eqref{eq:BSEeq}. For all of the parameter sets we have studied, the leading eigenfunction $\phi_0(k) \equiv \phi_{s^\pm}(k)$ has $s^\pm$ symmetry, i.e it changes sign between $k_z=0$ and $\pi$ but does not change sign as a function of $k_x$ and $k_y$~\cite{Maier2011, Maier2019}. To demonstrate this, Fig.~\ref{fig:kavEig} plots the frequency dependence of the leading eigenfunction $\phi_{s^\pm}(k)$ at $\beta = 5/t$ for parameters with [panel (a)] and without [panel (b)] an incipient band. In both cases, $\phi_{s^\pm}(k)$ has been integrated over ${\bf k}_\parallel = (k_x,k_y)$ (we find that their $k_x$ and $k_y$ dependence is very weak) and results are shown for the $k_z = 0$ and $k_z = \pi$ bands. The integrated eigenfunctions show the expected sign-change for an $s^\pm$ gap symmetry. Moreover, we observe an anisotropy in that $|\phi_{s^\pm}(k)|$ is generally smaller on the higher energy $k_z = 0$ band, and this asymmetry becomes more pronounced in the incipient band case. 

\begin{figure}[t]
\centering
\includegraphics[scale=0.4]{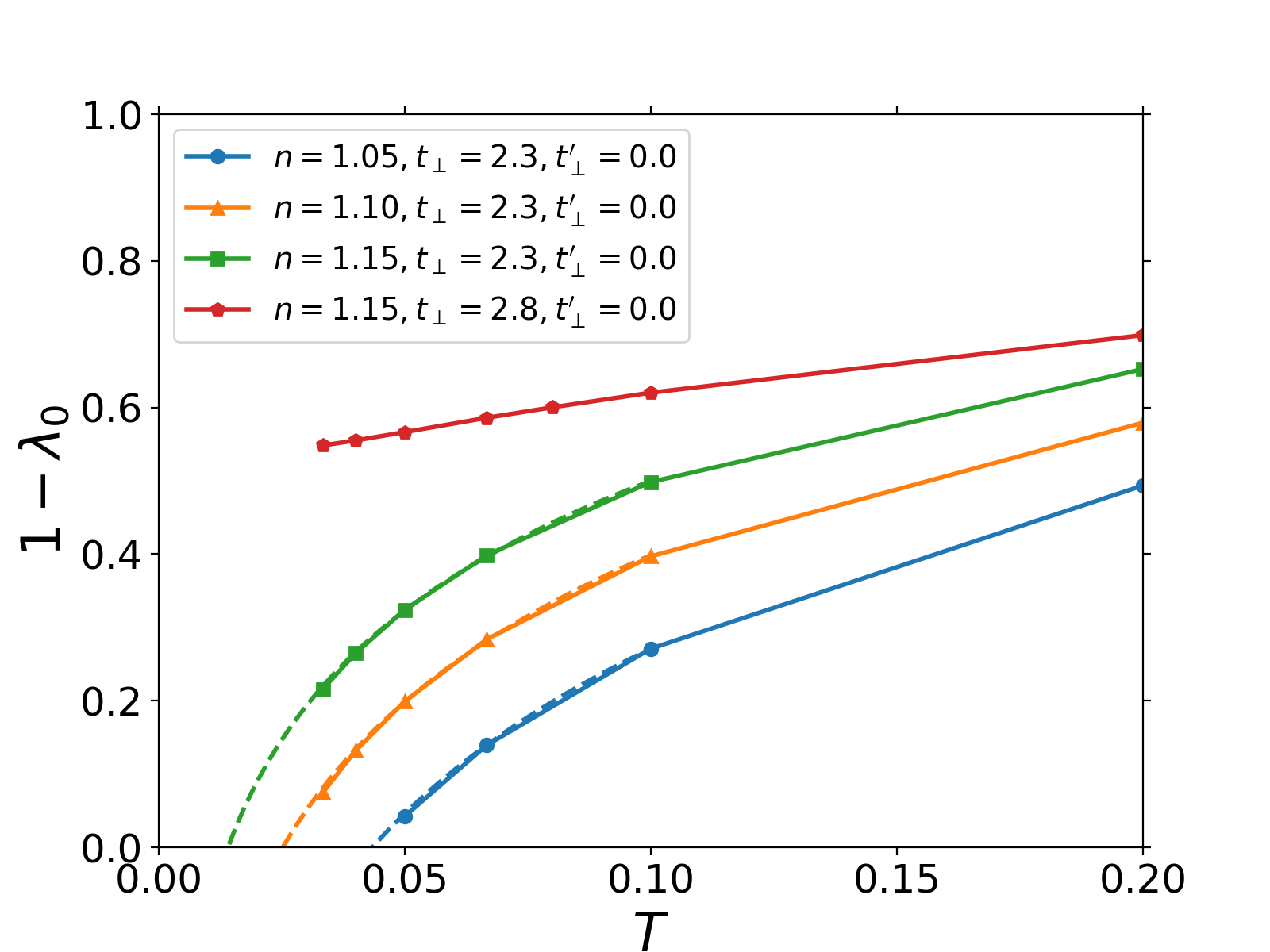}
\caption{The temperature dependence of the leading eigenvalue of the Bethe-Salpeter equation $1-\lambda_{s^\pm}(T)$ for different parameter sets. The superconducting $T_c$ can be estimated from this data by extrapolating $1-\lambda_{s^\pm} = 0$, as indicated by the dashed lines. Results are shown for several cases where two-bands cross the Fermi level [$\{ n, t_\perp, t_\perp^\prime \} = \{1.05, 2.3t, 0\}$, $\{1.10, 2.3t, 0\}$, and $\{1.15, 2.3t, 0\}$, which have $T_c \approx 0.043t$, $0.025t$, and $0.014t$, respectively. Conversely, no superconducting transition is found for the case with an incipient band [$\{1.15, 2.8t, 0\}$].}
\label{fig:lambdabulkinc}
\end{figure}

In unconventional multi-band superconductors like the Fe-based superconductors, the sign change in the gap function minimizes the effects of the Hubbard repulsion. Any anisotropy in 
$\phi_{s^\pm}(k)$ results from a combination of orbital character, Fermi surface size, and variations in the Fermi velocity $v_\mathrm{F}$ around the Fermi surface sheets \cite{MaierPRB2009, ZhangPRB2009}. In our simplified case, the orbital character of the bands is equally divided between the two layers and independent of ${\bf k}_\parallel$. The anisotropy is, therefore, driven by differences in the Fermi surface size and $v_\mathrm{F}$. As the top of the $k_z=\pi$ band moves down in energy and closer to the Fermi level, before it eventually becomes incipient, the asymmetry in the electronic structure and hence in the eigenfunction gets more pronounced. In the two band case [Fig.~\ref{fig:kavEig}(b)], the frequency dependence is nearly compensated between the two bands and approaches zero as $|\omega_n|\rightarrow\infty$. However, in the incipient case, the eigenfunctions of both bands change sign at a finite frequency and approach a finite value for $|\omega_n|\rightarrow\infty$. When the hole band has moved below the Fermi level, its gap function can no longer sufficiently contribute to the cancellation of $U$ at low energies, and the gap function on the remaining electron band has to adjust by generating a sign change in the frequency dependence. This behavior is analogous to conventional superconductors with a large Coulomb pseudopotential~\cite{PhysRevLett.100.237001}, where the high-frequency gap function $\Delta(\omega)$ approaches a negative constant as $\omega\rightarrow W$ to overcome the instantaneous Coulomb repulsion. In the present case, the $s$-wave structure of the gap with a sign change in its frequency dependence reflects an effective pairing interaction for the states on the electron band that is attractive at low frequencies, and changes sign at higher frequencies \cite{Maier2019}. This situation is just like in the electron-phonon case, except that it arises from spin fluctuations here. This structure may be understood in terms of an effective low-energy interaction for the electron band states that arises from virtual pair scattering to the hole states on the submerged band. The effective repulsive nature at high energies then arises from the Coulomb interaction \cite{Maier2019}.

Having established the dominant pairing symmetry, we now turn to the transition temperature $T_c$. Fig.~\ref{fig:lambdabulkinc} presents $1-\lambda_{s^\pm}(T)$ as a function of temperature for different values of the total density $n$ and $t_\perp$. The parameter sets $\{n, t^{\phantom\prime}_\perp, t^\prime_\perp\} = \{1.05, 2.3t, 0\}$, $\{1.10, 2.3t, 0\}$ and $\{1.15, 2.3t, 0\}$ all produce dressed spectral functions with two disconnected Fermi surface sheets similar to the one shown in the left panel of the Fig.~\ref{fig:Akwbulkinc}. Conversely, the parameters $\{1.15, 2.8t, 0\}$ produce an incipient band, as already established in the right panel of Fig.~\ref{fig:Akwbulkinc}. 

\begin{figure}[t]
\centering
\vspace{-0.75cm}
\includegraphics[scale=0.42]{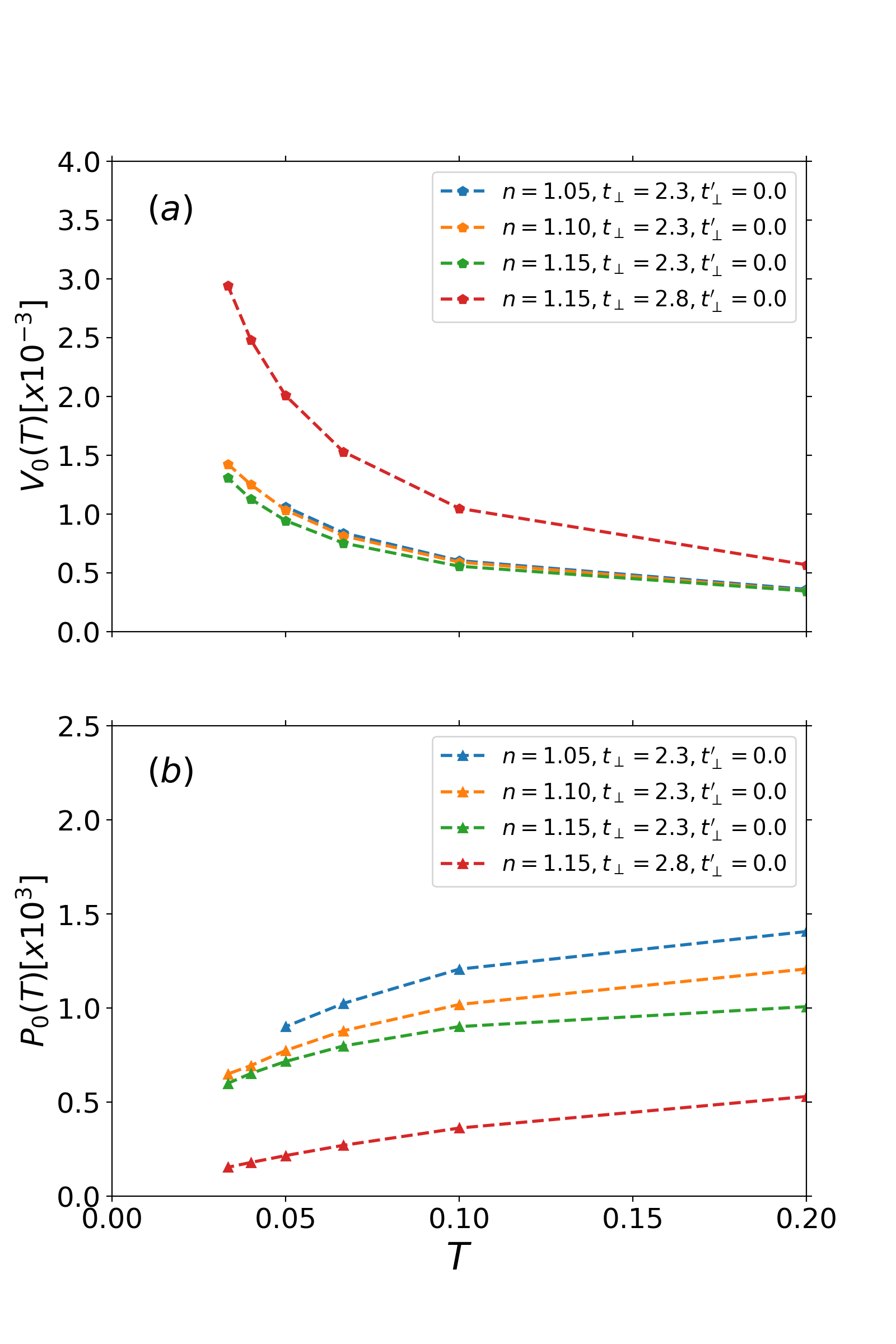}
\vspace{-0.75cm}
\caption{The temperature dependence of (a) the interaction strength $V_{s^\pm}(T)$ 
and (b) dressed pair-field susceptibility $P_{s^\pm}(T)$ for different parameter sets. Results are shown for  $\{n,t_\perp,t_\perp^\prime\} = \{1.05, 2.3t, 0\}$, $\{1.10,2.3t,0\}$, $\{1.15,2.3t,0\}$ (all with two bands crossing $E_\mathrm{F}$), and $\{1.15,2.8t,0\}$ (an incipient band case). All results were obtained on an $4\times 4\times 2$ cluster with $U = 6t$. 
}
\label{fig:separable_VandP}
\end{figure}

We can estimate the superconducting $T_c$ for these parameter sets by extrapolating the temperature dependence of $1-\lambda_{s^\pm}(T)$ to zero. Here, we fit the low-temperature data with a function of the form $f(T) = A\log(T/T_c)$, which is then used to perform the extrapolation.\footnote{This functional form is motivated by the BCS Cooper log singularity. However, as we will be shown in Fig.~\ref{fig:separable_VandP}, such a divergence does not appear in the intrinsic pair-field susceptibility. Nevertheless, this choice describes the behavior of $1-\lambda_{s^\pm}(T)$ for $T\rightarrow T_c$ well.} Using this approach, we obtain finite transition temperatures for the cases with two Fermi surfaces, with $T_c \approx 0.043t$, $0.025t$, and $0.014t$ for $\{n, t_\perp, t^\prime_\perp\} = \{1.05, 2.3t, 0\}$, $\{1.10, 2.3t, 0\}$, and $\{1.15, 2.3t, 0\}$, respectively. (Again, the leading pairing instability in all three cases corresponds to an $s^\pm$ state.) Thus, $T_c$ is reduced as the system is progressively electron-doped and the hole-like band at $\Gamma$ approaches $E_\mathrm{F}$. For the parameter set resulting in an incipient band, we find that the leading instability is still of an $s^\pm$ symmetry, consistent with the predictions of several prior works~\cite{BangNJP2014, LinscheidPRL2016, Mishra2016, KurokiFlex2020, Maier2019, RademakerEnhanced2021}. However, we also observe a clear suppression of the pairing correlations and no clear signs of a superconducting transition down to the lowest simulated temperatures ($T = 0.033$), as shown in Fig.~\ref{fig:lambdabulkinc}. 

The results shown in Fig.~\ref{fig:Akwbulkinc} suggest that superconductivity is suppressed in the incipient band case. To better understand the mechanism behind this behavior, we calculated the noninteracting dressed (intrinsic) pair-field susceptibility $P_{s^\pm}(T)$ [Eq.~(\ref{eq:P0})] and the $s^\pm$ interaction strength $V_{s^\pm}(T)$ [Eq.~\ref{eq:V}]. Fig.~\ref{fig:separable_VandP} presents the temperature evolution of both quantities for the same parameter sets used in Fig.~\ref{fig:lambdabulkinc}. 
In general, these quantities display a very similar temperature dependence as observed in previous work for the weakly doped single-band Hubbard model in the pseudogap regime \cite{Maier2016}. In a typical BCS superconductor, the superconducting instability is driven by the Cooper log instability of the noninteracting pair-field susceptibility $P(T)$. In contrast to this behavior, here, $P_{s^\pm}(T)$ instead decreases as the temperature is lowered, and the instability arises from an increase in the pairing interaction $V_{s^\pm}(T)$ in all cases.  Moreover, we see competing tendencies as the hole band becomes incipient: the systems with two bands crossing the Fermi level generally have larger $P_{s^\pm}$ compared to the one with an incipient band. In contrast, the pairing interaction $V_{s^\pm}$ is the largest for the incipient band case. Matsumoto {\it et al.} gave a plausible argument for this behavior  \cite{KurokiFlex2020}. They proposed that spin fluctuation spectral weight is transferred from low, pair-breaking energies to more optimal higher energies as the band becomes incipient. However, we find that the removal of Fermi energy states as the hole band drops below $E_\mathrm{F}$ significantly decreases the intrinsic pair-field susceptibility $P_{s^\pm}$. This loss counteracts any increase in $V_{s^\pm}$, resulting in a net decrease in $T_c$.  

\begin{figure}[t]
\centering
\includegraphics[scale=0.4]{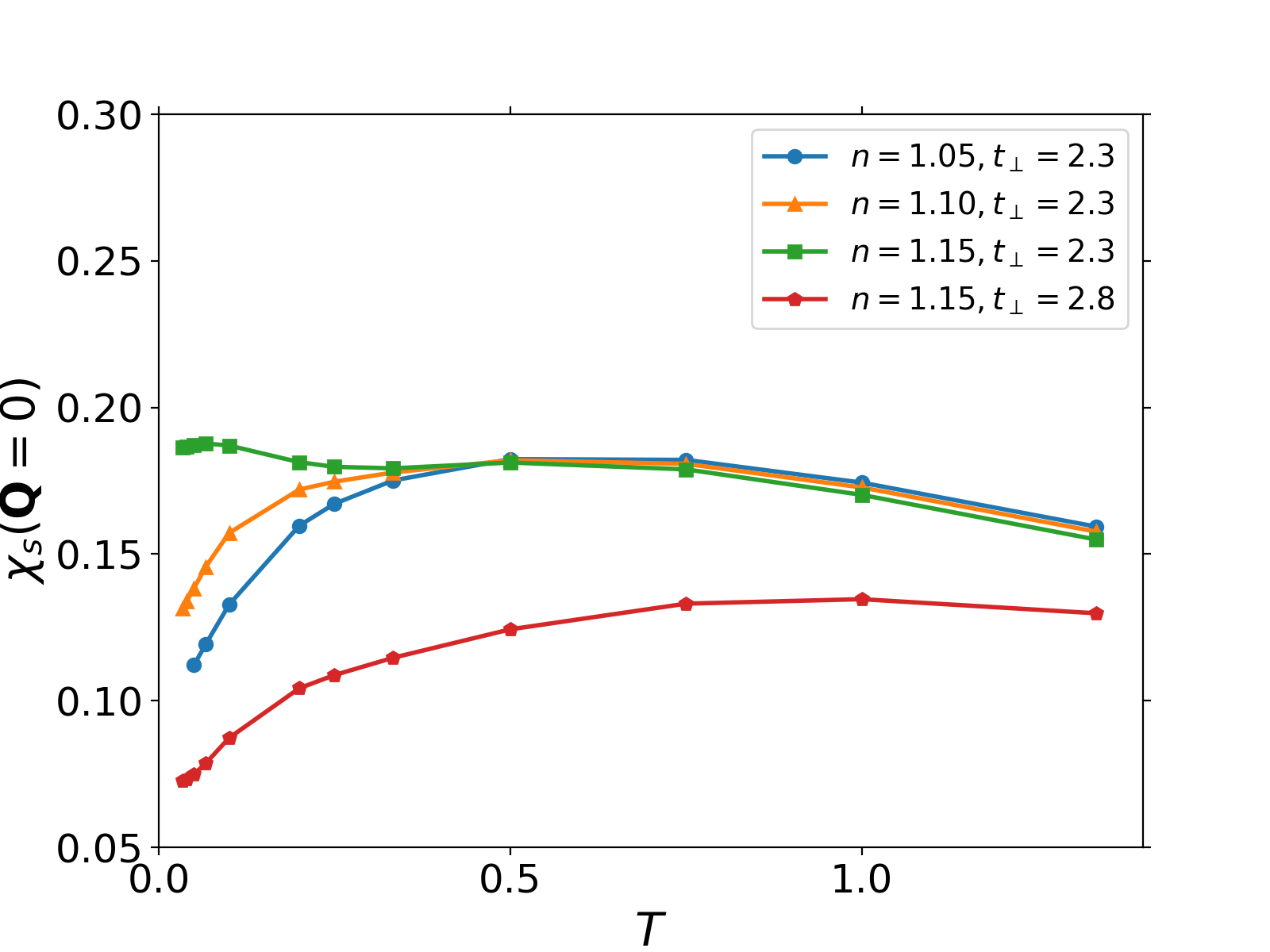}
\caption{The static spin susceptibility $\chi_s({\bf Q})$ measured at momentum scattering vector $\mathbf{Q}=(0,0,0)$ for the same parameter sets used in Fig.\ref{fig:separable_VandP}. The existence of a local maximum in the temperature dependence of $\chi_s(\mathbf{Q}=(0,0,0))$ at a temperature $T^*$ (indicated by the arrows) together with the decrease in $P_{s^\pm}(T)$ as $T$ is lowered  signals the formation of a pseudogap in the density of states.}
\label{fig:spinsusQ0}
\end{figure}

Moreover, we find that $P_{s^\pm}$ decreases with increasing doping away from the half-filled Mott insulating state, even for cases where both bands cross the Fermi level. This is counter to what is observed for the single-band Hubbard model, where the $d$-wave intrinsic pair-field susceptibility is found to increase with increasing doping \cite{Maier2016, Maier2018}. The behavior we observe here arises from the projection of the pair propagator onto the leading $s^\pm$ eigenfunction $\phi_{s^\pm}$ in Eq.~(\ref{eq:P0}), and its asymmetry between the two bands. As the doping increases, the asymmetry in $\phi_{s^\pm}$ results in an increased emphasis of the lower energy $k_z = \pi$ band in this projection, which contributes less to $P_{s^\pm}$, resulting in its decrease.

As noted, the reduction in $P_{s^\pm}(T)$ with decreasing temperature is similar to the pseudogap behavior observed before in the single-band Hubbard model \cite{Maier2016}, where Fermi surface states are removed in the anti-nodal region. The situation for the incipient case is analogous, where the states associated with the $k_z = \pi$ band are submerged below the Fermi energy. While the electron Fermi surface pocket associated with the $k_z = 0$ band remains intact, the $s^\pm$ pairing state necessarily involves the states on the submerged band \cite{Maier2019}. The $s^\pm$ is therefore affected by the absence of these states, here manifested as a decrease in $P_{s^\pm}(T)$. But what is perhaps more remarkable is that the cases with two bands crossing the Fermi energy also show a decrease of $P_{s^\pm}(T)$ as the temperature is lowered, indicating that low-energy states are partially removed. To confirm this, we calculated the bulk spin susceptibility $\chi_s(\mathbf{Q}=0, \omega=0)$ as a function of temperature, as shown in Fig.~\ref{fig:spinsusQ0}. For all the systems, the spin susceptibility reaches a maximum value at an intermediate $T = T^*$ before decreasing as the temperature is lowered further. This behavior reflects the removal of low-energy spin excitations, as previously observed in DCA calculations for the bilayer Hubbard model in Ref.~\citenum{Maier2011}. As discussed in that work, this behavior may be attributed to the inter-layer spin fluctuations becoming gapped at low temperatures due to the formation of inter-layer singlets for large $t_\perp$. Together with the decrease seen in $P_{s^\pm}(T)$ as the temperature is lowered, it signals the opening of a pseudogap in the density of states at $T^*$, as observed previously for the single-band Hubbard model in the large $U$ limit \cite{Maier2016, Maier2019_2}. The larger doping case with $\langle n\rangle=1.15$ and $t_\perp=2.3t$, where $P_{s^\pm}(T)$ continues to rise at lower temperatures, is somewhat peculiar. Here, we believe that the larger electron doping suppresses the singlet formation and pseudogap behavior. 

\subsection{Results with next-nearest-neighbor interlayer hopping}
In the previous section, we set the inter-layer next-nearest-neighbor hopping $t_\perp^{\prime}$ to zero and found 
no indications that a system with an incipient band had a higher $T_c$ compared to the case where two bands cross $E_\mathrm{F}$. 
Ref.~\cite{KurokiFlex2020}, however, studied Hubbard ladders and bilayers with a non-zero $t_\perp^{\prime}$ using FLEX and VMC and found such hopping could enhance superconductivity when the top of the hole band is near $E_\mathrm{F}$. Motivated by this, we performed additional simulations with  $t_\perp^{\prime} \ne 0$. To begin, we fixed $t_\perp =1.8t$ and $U=6t$, and again considered $4\times 4\times 2$ clusters while varying $t_\perp^{\prime}$ and densities $n=1.05$, $1.10$, and $1.15$. From the technical point of view, as we increase $t_\perp^{\prime}$, the average value of the Fermion sign obtained by our QMC solver decreases abruptly for these parameters. For this reason, we are restricted to comparatively higher temperatures than in the previous case. 

\begin{figure}[t]
\centering
\includegraphics[scale=0.4]{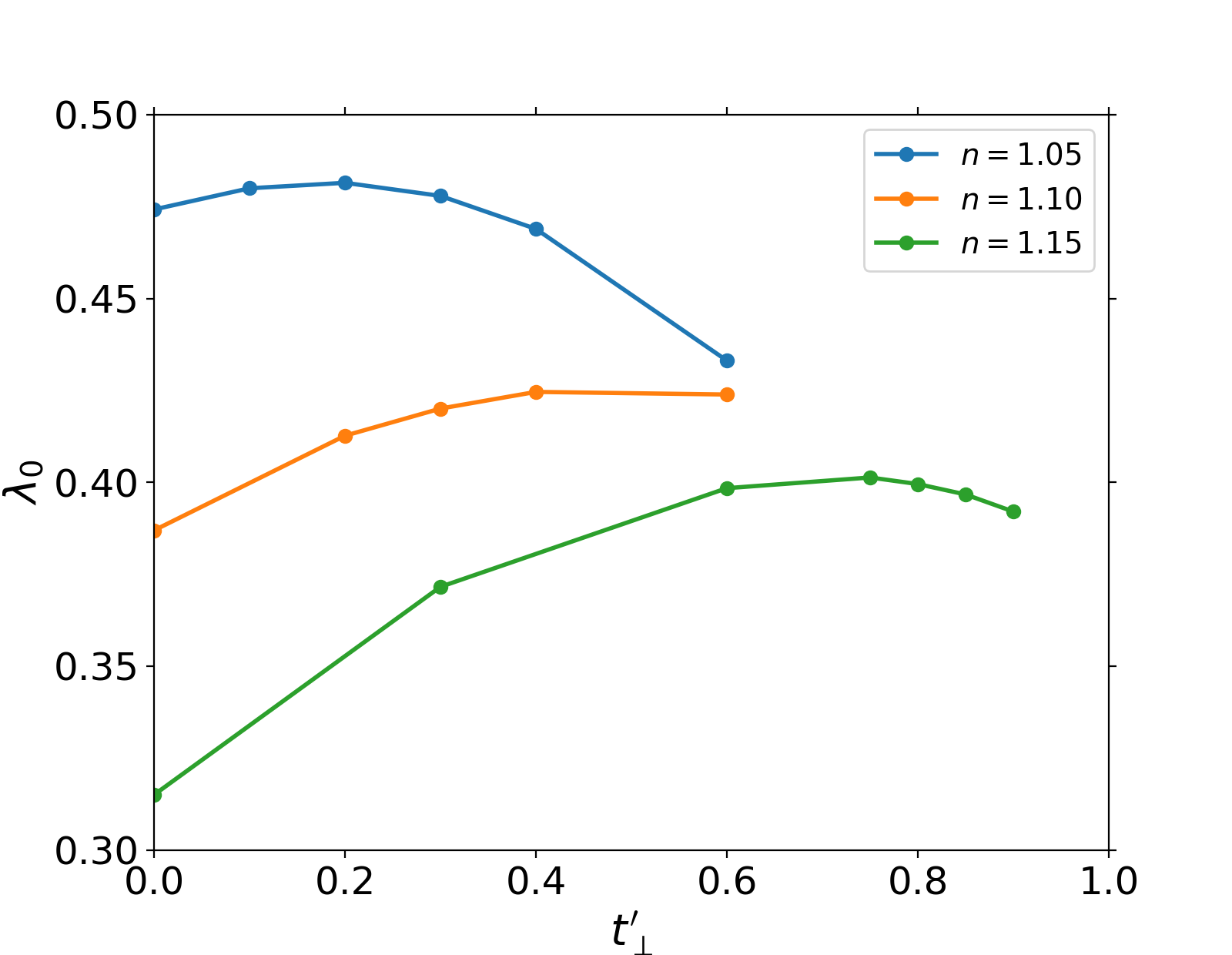}
\caption{The dependence of the largest ($s^\pm$) eigenvalue $\lambda_0$ of the BSE equation as a function of $t_{\perp}^{\prime}$ and  different densities. All results were obtained for $t_{\perp}=1.8t$, $\beta=5/t$, $U=6t$ and on an $4\times 4\times 2$ cluster.}
\label{fig:lambdatperpprime}
\end{figure}

\begin{figure*}[t]
\centering
\includegraphics[width=\textwidth]{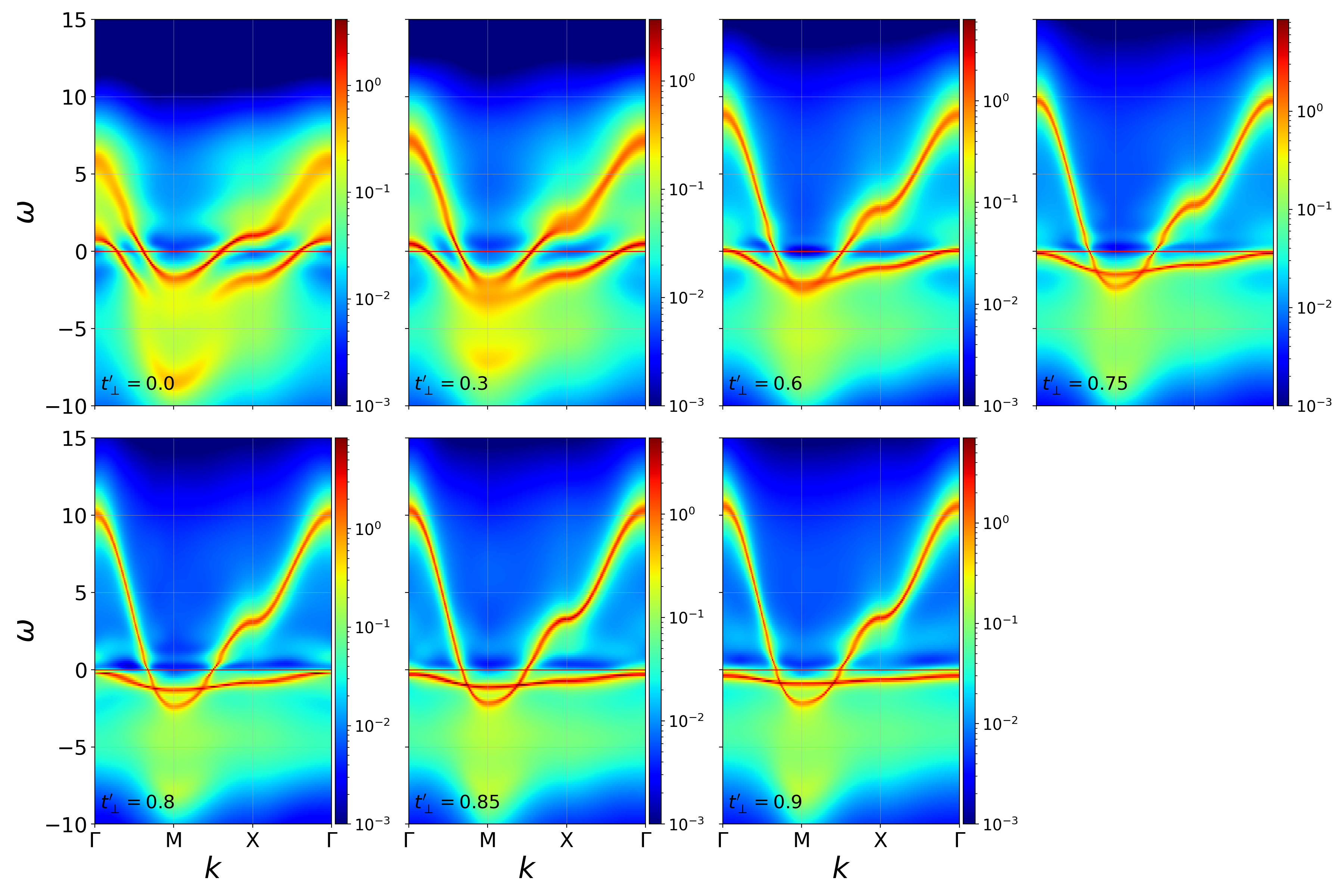}
\caption{The single-particle spectral function $A({\bf k},\omega)$ and different values of $t_{\perp}^{\prime}$, as indicated in the lower left part of each panel. All spectra were computed for $t_{\perp}=1.8t$, $U=6t$, and $\beta=5/t$ and on a $4\times 4\times 2$ cluster with density $n=1.15$.}
\label{fig:Akwtperprime}
\end{figure*}

Figure~\ref{fig:lambdatperpprime} plots the leading eigenvalue of the BSE as a function of $t_\perp^{\prime}$ at a fixed inverse temperature of $\beta=5/t$. As with the $t^\prime_\perp = 0$ case, we find that the leading instability has an $s^\pm$ symmetry, consistent with the prior FLEX results \cite{KurokiFlex2020}. We also find that the strength of the pairing correlations reaches a maximum value at an optimal density-dependent $t_\perp^{\prime}$ value before decreasing gradually as $t_\perp^{\prime}$ is increased further. We also note that $\lambda_{s^\pm}$ increases more rapidly as a function of $t^\prime_\perp$ in the systems with larger electron densities. Our observations support the findings of Ref.~\citenum{KurokiVMC2020} who also observed increased pairing correlations with larger $t^\prime_\perp$ and a fixed temperature and density. 

Next, we examine the effect of $t_\perp^{\prime}$ on the dressed spectral function of the system with parameters $n=1.15$, $t_\perp=1.8t$, $U=6t$ and $\beta=5/t$, which produces the largest increase of the pairing correlations with increasing $t_\perp^{\prime}$ (see Fig.~\ref{fig:lambdatperpprime}).  Fig.~\ref{fig:Akwtperprime} shows the single-particle spectral function for $t_\perp^{\prime} \in [0,0.9]$, as indicated in each panel. In general, we observe that the hole-like band at $\Gamma$ flattens with increasing $t^\prime_\perp$ and is pushed towards lower energies, consistent with the expectation from the $U=0$ picture (see Fig.~\ref{fig:bilayer_sketch}). The system undergoes a Lifshitz transition and forms an incipient band for $0.6 < t_\perp^{\prime} <0.75t$, as shown in the third column of the first row. Comparing with Fig.~\ref{fig:lambdatperpprime}, we see that the maximum in the $s^\pm$ eigenvalue for this parameter set at $t_\perp^\prime \approx 0.75t$ corresponds to a system where the hole band is incipient, i.e., slightly below the Fermi level. This $t_\perp^\prime$ behavior of the $s^\pm$ eigenvalue at a fixed temperature is, therefore, consistent with the observation of Matsumoto {\it et al.}~\cite{KurokiFlex2020}, who also found that the eigenvalue is strongly enhanced when the top of the hole band is driven close to $E_\mathrm{F}$. 

\begin{figure}
\centering
\includegraphics[scale=0.4]{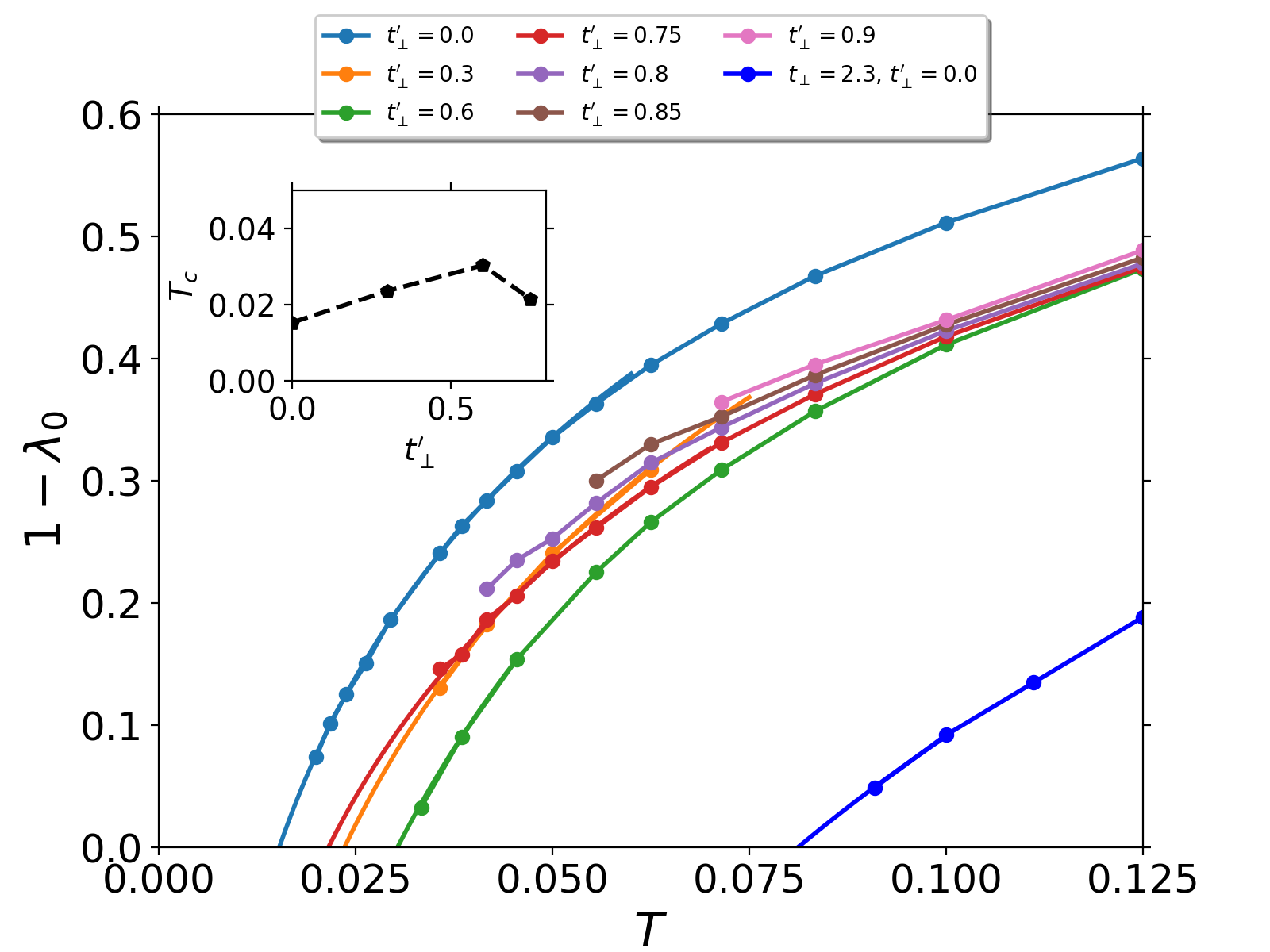}
\caption{The temperature dependence of the leading eigenvalue of the BSE equation of the bilayer Hubbard model for  $t_{\perp}=1.8t$ and different values of $t_{\perp}^{\prime}$. The remaining model parameters are $U=6t$, $n=1.15$, and $N = 2\times 2 \times 2$. The inset shows the estimated transition temperature that could be reliably estimated by 
fitting $1-\lambda_0(T)$ with a function $f(T) = A\log(T/T_c)$. $2\times2\times2$ results are also plotted for $n=1.05$, $t_\perp=2.3$, $t_\perp^\prime=0$ (blue curve), to show that the case with two Fermi pockets has significantly higher $T_c$.}
\label{fig:lambdatperpprime2}
\end{figure}

Figure~\ref{fig:lambdatperpprime2} plots the temperature dependence of the corresponding leading eigenvalues of the BSE the same parameters shown in Fig.~\ref{fig:Akwtperprime}. Due to the severe sign problem present for these parameters, here, we consider $2\times 2\times 2$ clusters to access temperatures low enough that we can reliably estimate $T_c$ by extrapolation (see inset). The results reveal that $T_c$ is enhanced as $t_\perp^{\prime}$ increases up to $0.6$ but undergoes a sharp decrease at $t_\perp^{\prime} = 0.75$, when the hole band becomes incipient. Hence, unlike $\lambda_{s^\pm}$, which is maximized for the incipient band at $t_\perp^\prime=0.75$, $T_c$ is maximized for the $t_\perp^\prime=0.6$ case, just before the Lifshitz transition, and falls significantly when the band becomes incipient. We also note that the $t_\perp^\prime=0.6$ case still has significantly lower $T_c$ than the case in Fig.~\ref{fig:lambdabulkinc} with $\langle n\rangle=1.05$, $t^{\phantom\prime}_\perp=2.3$ and $t_\perp^\prime=0$, providing further evidence that two Fermi surfaces are better than one for maximizing $T_c$. 


Generally, we find that the extrapolated $T_c$ values are smaller in the incipient band case, and the system lacks a superconducting transition if the hole band is too far from the Fermi level. Our results show that a finite $t_\perp^{\prime}$ can indeed increase the pairing correlations at a fixed temperature as found by Kuroki \emph{et al.}. But this enhancement does not produce an increased $T_c$. For completeness, we also checked that our findings are robust against modest increases in the value of $U$, see appendix~\ref{sec:appendixU8}.

\subsection{Optimizing the superconducting transition temperature} 
From all of the parameter sets we have checked, the largest $T_c$ that we have been able to achieve in the model was $T_c \approx 0.043t$ (in the $4\times 4\times 2$ cluster) for $n=1.05$, $U=6t$, $t_\perp = 2.3t$, and $t_\perp^\prime = 0$. Given the results of the previous section, it is natural to wonder whether this transition temperature can be further optimized for some $t_\perp^\prime \ne 0$. 

\begin{figure}
\centering
\includegraphics[scale=0.43]{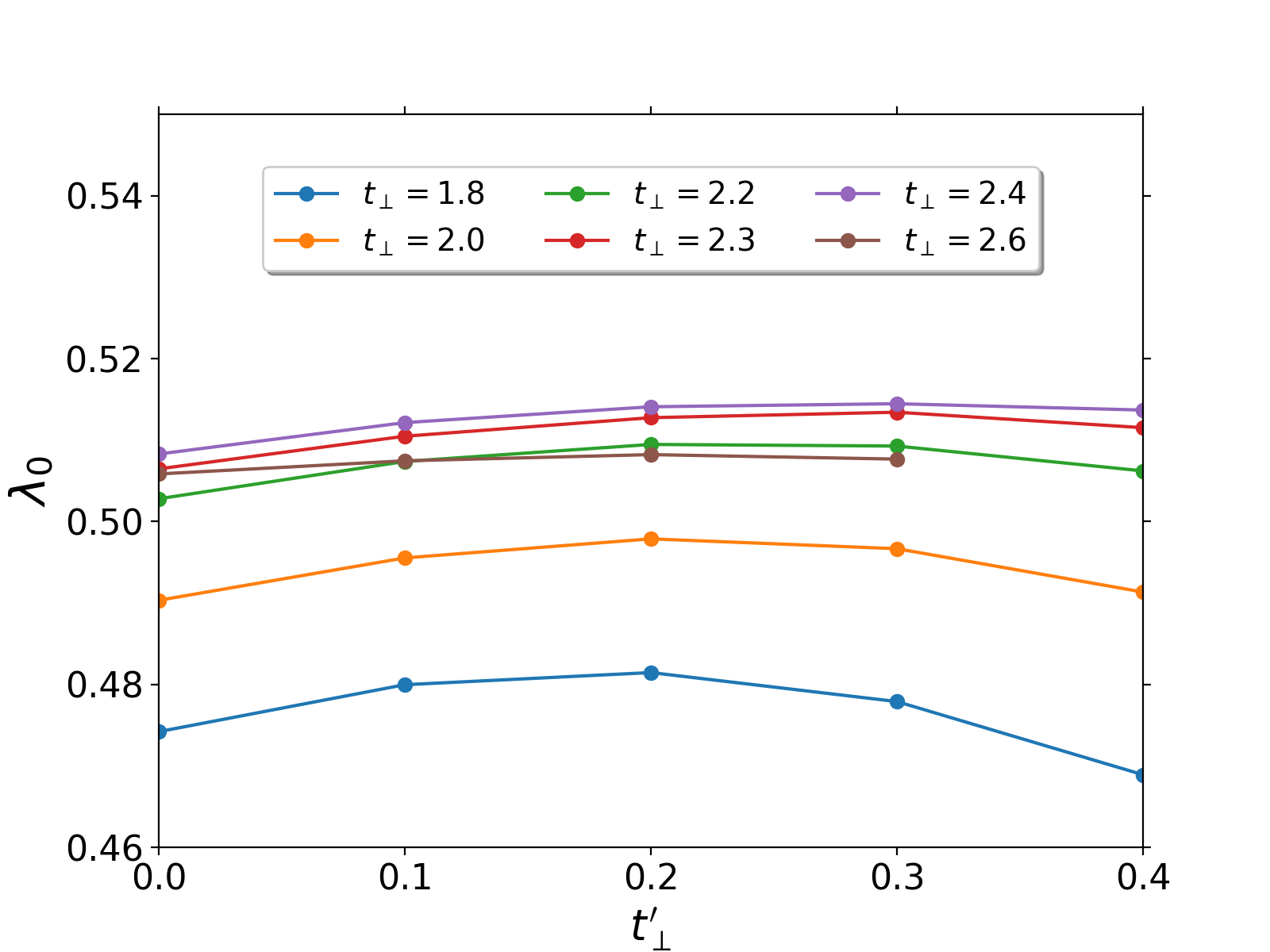}
\caption{The leading BSE eigenvalue of the bilayer Hubbard model for the parameters set with $n=1.05$ and $U=6t$ at $\beta= 5/t$ for different values of $t_{\perp}$ and $t_{\perp}^{\prime}$ calculated on a $4\times 4\times 2$ cluster. }
\label{fig:lambdatperpprime3}
\end{figure}

To answer this question, we considered the pairing correlations in the model with a fixed density $n=1.05$ and $t_\perp \in [1.8,2.6]$ such that the system has two well-defined Fermi surfaces. Fig.~\ref{fig:lambdatperpprime3} plots the  evolution of $\lambda_{s^\pm}$ for different values of $t_{\perp}^{\prime}$ at a fixed temperature $\beta= 5/t$. Unlike the case with $n=1.15$ (Fig.~\ref{fig:lambdatperpprime}), here we find that $\lambda_{s^\pm}$ is only weakly dependent on $t_{\perp}^{\prime}$ for all of the $t_{\perp}$ values we have considered. 
Nevertheless, $\lambda_{s^\pm}$ does exhibit a local maximum for some optimal value of $t_\perp^\prime$ that depends on the value of $t_\perp$. For small $t_\perp^\prime \approx 0.2t$, we also find that $\lambda_{s^\pm}$ has an optimal $t_{\perp}$ of about $2.4t$. 

\begin{figure}
\centering
\includegraphics[width=\columnwidth]{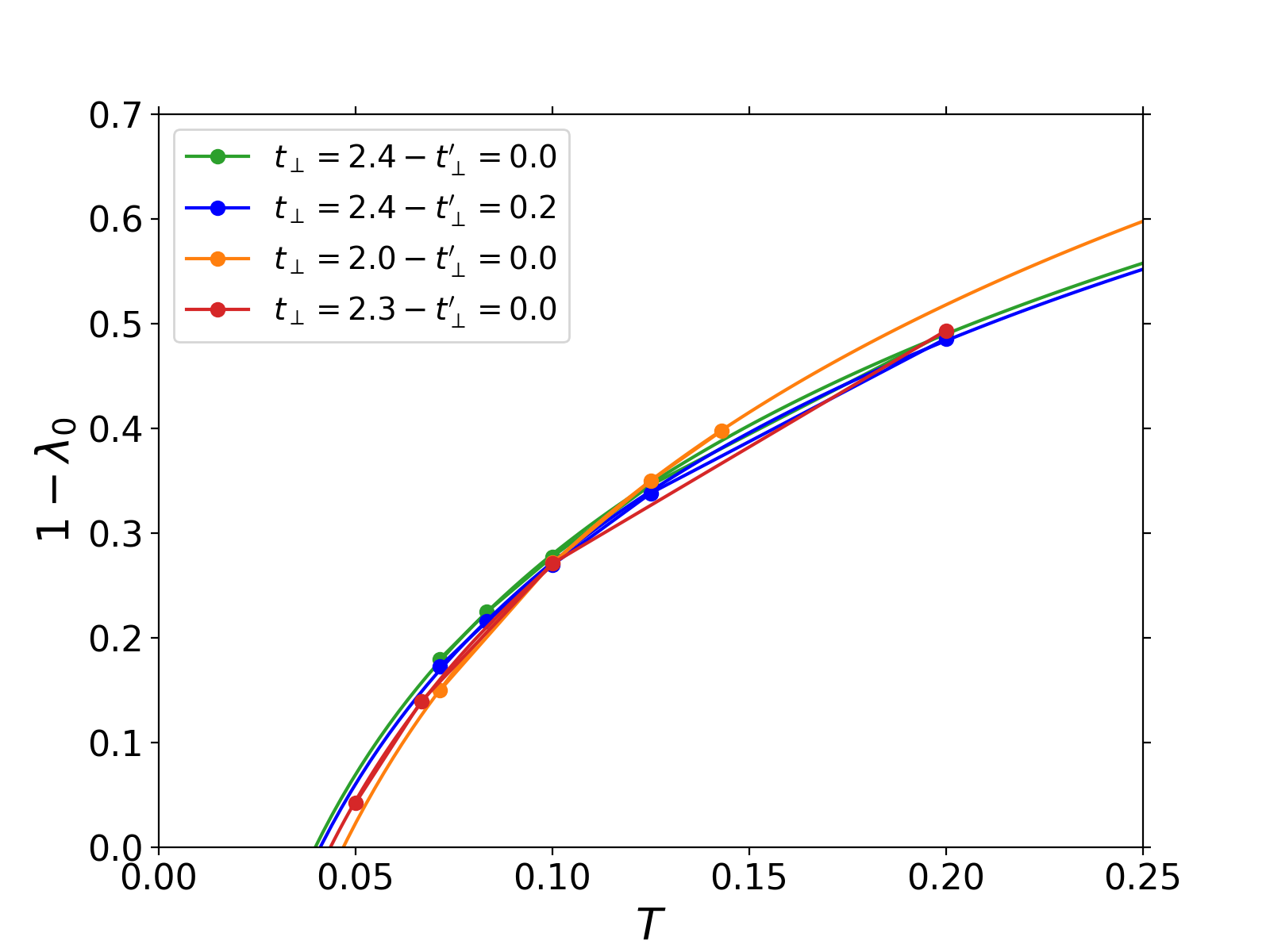}
\caption{The temperature dependence of the leading eigenvalue of the BSE equation of the bilayer Hubbard model for the parameters set with $n=1.05$ and $U=6t$ for different values of $t_{\perp}$ and $t_{\perp}^{\prime}$ calculated on a $4\times 4\times 2$ cluster.  }
\label{fig:lambdavsT_n0.95}
\end{figure}

Turning to the transition temperature, in Fig.~\ref{fig:lambdavsT_n0.95} we analyze the temperature dependence of $\lambda_{s^\pm}$ for the cases $\{n, t^{\phantom\prime}_\perp, t^\prime_\perp\}$  = $\{1.05,2.4t,0\}$, $\{1.05,2.4t,0.2t\}$, $\{1.05,2t,0\}$. 
By extrapolating the values of $1-\lambda_{s^\pm}$, we obtain $T_c \approx 0.0398t$,  $0.041t$, $0.047t$, respectively. Thus, the largest $\lambda_{s^\pm}$ value at $\beta = 5/t$ occurs for the parameters $n=1.05$, $t_\perp = 2.4t$, $t_\perp^\prime=0.2t$, but the highest $T_c$ is achieved for the parameters $\{1.05,2t,0\}$. Based on these results, we conclude that adding a finite $t^{\phantom\prime}_{\perp}$ or $t_{\perp}^{\prime}$ might increase the superconducting correlations at a finite temperature, but this might not mean that the resulting $T_c$ will also increase. One should always check the temperature dependence and extrapolate for the $T_c$.

\begin{figure*}
\centering
\includegraphics[width=\textwidth]{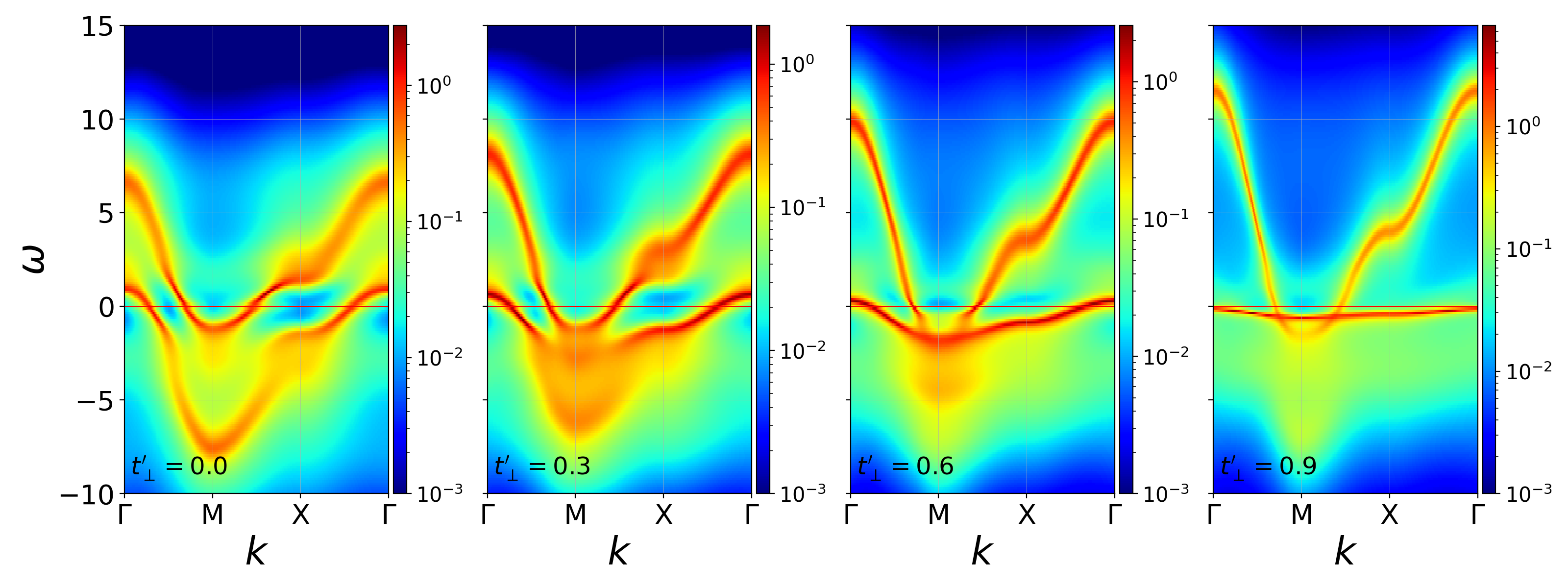}
\caption{The single-particle spectral function $A({\bf k},\omega)$ and different values of $t_{\perp}^{\prime}$, as indicated in the lower left part of each panel. All spectra were computed for $t_{\perp}=2.0t$, $U=6t$, and $\beta=5/t$ and on a $4\times 4\times 2$ cluster with density $n=1.05$.}
\label{fig:Akwtperprime_n0.95}
\end{figure*}

Finally, Fig.~\ref{fig:Akwtperprime_n0.95} shows how the spectral function $A({\bf k},\omega)$ evolves for $\{1.05,2t,t^\prime_\perp\}$ as a function of $t_\perp^\prime$, for which we obtained the highest $T_c$. Here, we see that the introduction of $t_\perp^\prime$ produces a significant change in the band structure. At the same time, we know from Figs.~\ref{fig:lambdatperpprime} and \ref{fig:lambdatperpprime3} that $t_{\perp}^{\prime}$ does very little to $\lambda_{s^\pm}$ for this case. 
 
\section{Discussion and Conclusions}\label{sec:discussion} 
We have studied the electronic structure and pairing correlations of the bilayer Hubbard model in the strong coupling $(U = 6t)$ and large interlayer hopping $t_\perp$ limit using the DCA with a nonperturbative QMC cluster solver. By adjusting the value of the interlayer hopping integral, we can tune the system through a Lifshitz transition whereby the system changes from having an electron- and hole-like band crossing the Fermi level at $M$ and $\Gamma$, respectively, to one with one band crossing $E_\mathrm{F}$ at $M$ and an incipient hole-like band at $\Gamma$. In all cases, the solutions to the BSE indicate that the leading superconducting instability corresponds to an $s_\pm$ symmetry, 
where the sign of the order parameter changes between the two bands. Moreover, we find that the pairing correlations can be enhanced at a fixed temperature by narrowing the bandwidth of a nearly incipient band such that additional density of states is concentrated near $E_\mathrm{F}$. Both of these findings agree with prior studies \cite{BangNJP2014, ChenPRB2015, Mishra2016, KurokiFlex2020, LinscheidPRL2016, RademakerEnhanced2021}. 
Contrary to these works, however, we find that $T_c$ is always reduced once one of the 
bands is made incipient.  For example, in the case of the two-band system, we can extract finite values for the superconducting $T_c$ from the temperature dependence of the leading eigenvalue of the BSE. In the incipient band case, however, we find that $T_c$ rapidly decreases as the hole band is submerged. Our results provide compelling evidence that superconductivity in the bilayer Hubbard model, in the strong coupling and large interlayer hopping $t_\perp$ limit, is optimized when there are well-defined hole- and electron-like Fermi surface sheets. Subsequent electron doping then reduces $T_c$ in this model, provided all other factors remain fixed. 

We stress that one should be careful in extrapolating our results to systems like the FeSe intercalates or FeSe/STO monolayers. The bilayer Hubbard model has been widely studied as a simple toy model for extrapolating between strongly correlated models with cuprate-, iron-pnictide-, and (incipient band) FeSe/STO-like band structures \cite{Maier2011, Mishra2016, Kuroki2008,  RademakerEnhanced2021, PelliciariRIXS2020}. However, in reality, the latter systems are multi-band materials whose band structures at the Fermi level are predominantly Fe $3d$ character with several partially filled bands crossing the Fermi level. The bilayer Hubbard model approximates these systems with a two-band model but with an on-site interorbital hopping $t_\perp$. For a non-zero $U \gg t_\perp$, the inter-orbital hopping will introduce a sizeable antiferromagnetic exchange coupling $J_\perp = 4t^2_\perp/U$ between the layers. The large interlayer AFM exchange may help explain why $T_c$ is reduced in the large $t_\perp$ limit, as it will tend to produce local disordered singlets within each unit cell. FLEX calculations cannot capture this effect. However, the Fe-based superconductors have large interatomic Hund's interactions, favoring ferromagnetic coupling between the two orbitals. 

\begin{acknowledgments}
This work was supported by the Scientific Discovery through Advanced Computing (SciDAC) program funded by the U.S. Department of Energy, Office of Science, Advanced Scientific Computing Research and Basic Energy Sciences, Division of Materials Sciences and Engineering. This research used resources of the Oak Ridge Leadership Computing Facility, which is a DOE Office of Science User Facility supported under Contract DE-AC05-00OR22725. This manuscript has been authored by UT-Battelle, LLC, under Contract No. DE-AC0500OR22725 with the U.S. Department of Energy. The United States Government retains and the publisher, by accepting the article for publication, acknowledges that the United States Government retains a non-exclusive, paid-up, irrevocable, world-wide license to publish or reproduce the published form of this manuscript, or allow others to do so, for the United States Government purposes. The Department of Energy will provide public access to these results of federally sponsored research in accordance with the DOE Public Access Plan (\url{http://energy.gov/downloads/doe-public-access-plan}). The DCA++ code used for this project can be obtained at \url{https://github.com/CompFUSE/DCA}. The data presented in this work can be obtained at \url{https://github.com/JohnstonResearchGroup/Karakuzu_etal_bilayer_2021}. 
\end{acknowledgments}

\appendix
\section{Results for $\mathbf{U = 8t}$}\label{sec:appendixU8}
\begin{figure}
\centering
\includegraphics[scale=0.34]{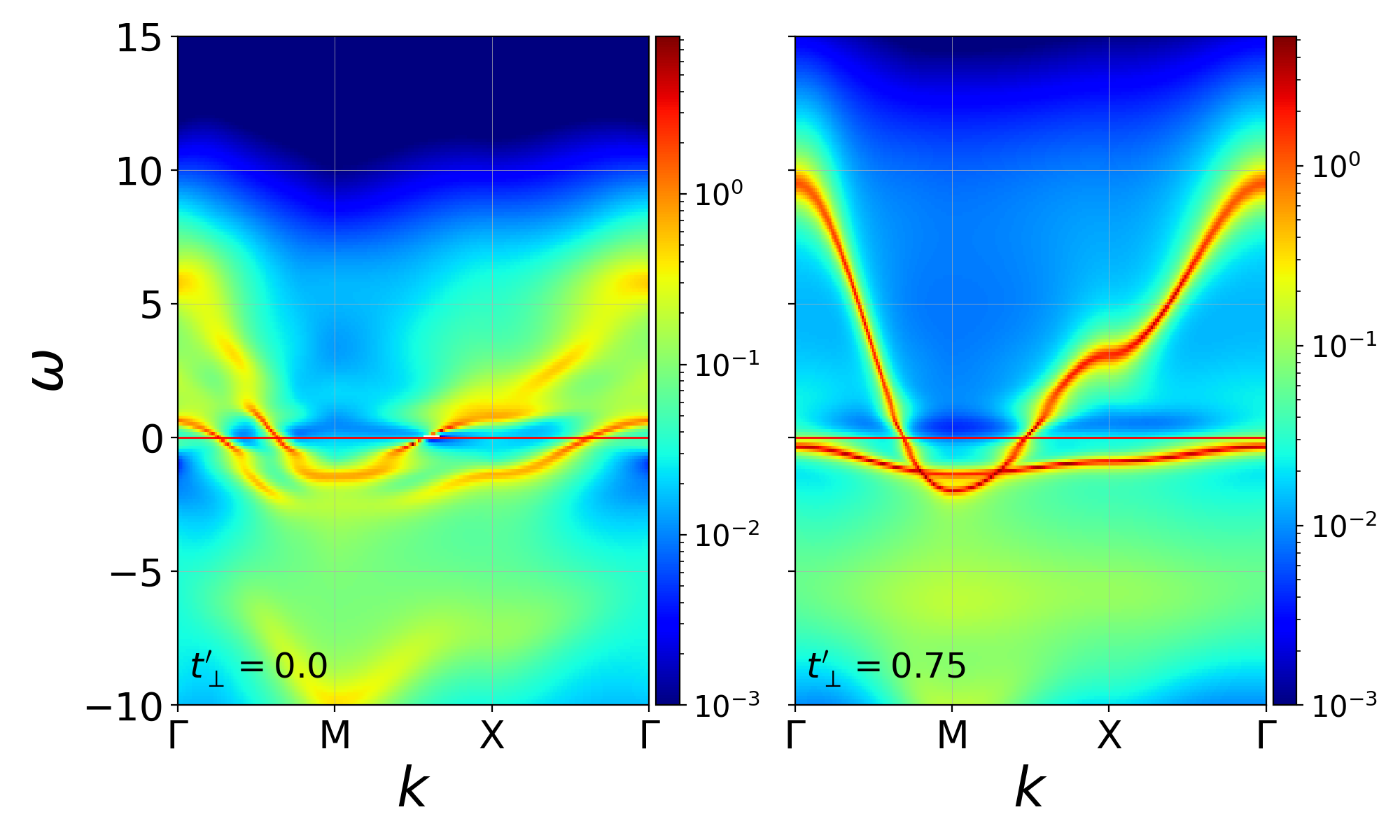}
\includegraphics[scale=0.4]{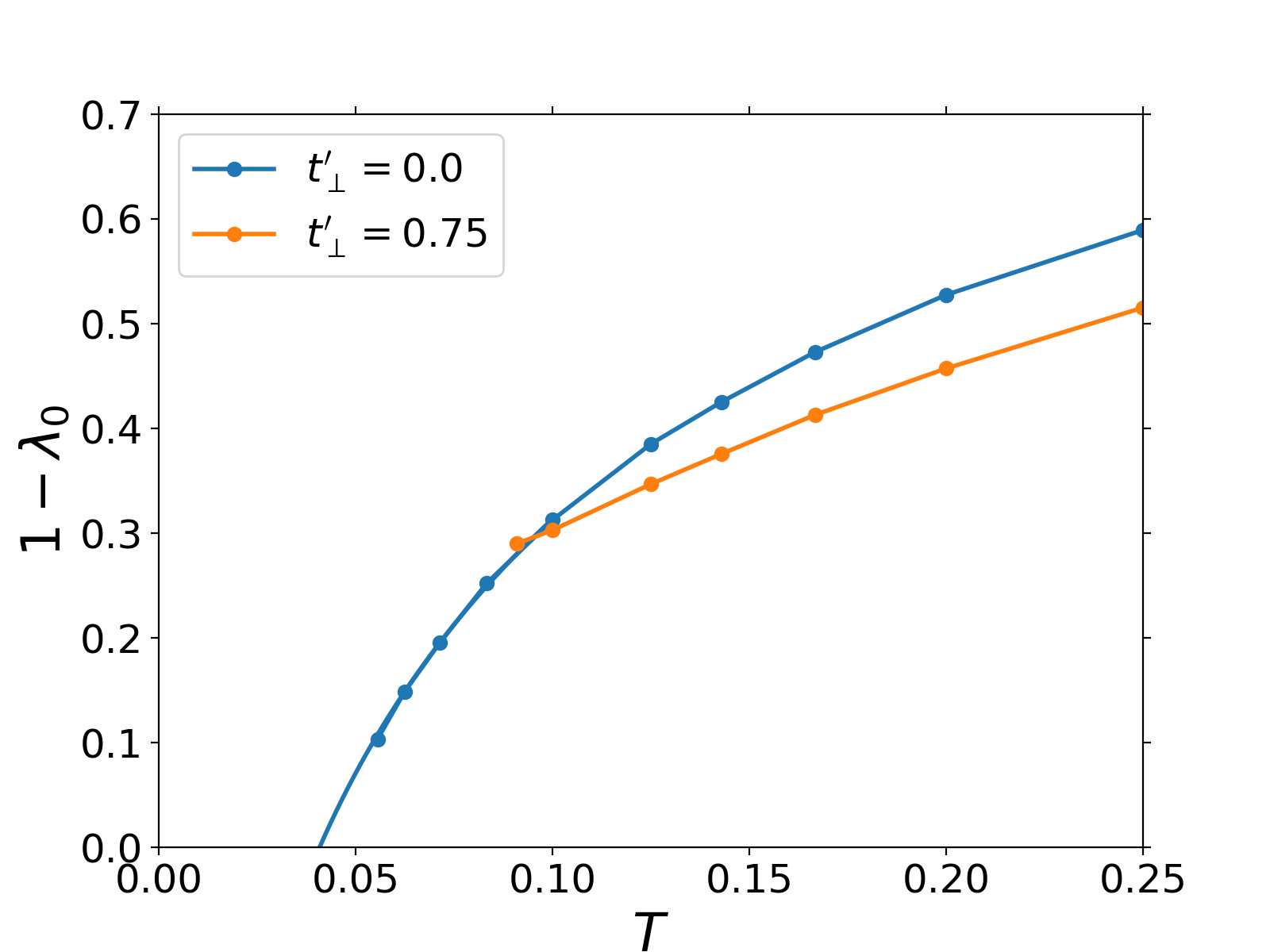}
\caption{Top row: $A({\bf k},\omega)$ for $t_{\perp}= 1.8$, $n = 1.15$, $U = 8t$, and $\beta = 5/t$, 
The left and right panels shows results for $t^\prime_\perp = 0$ and $0.75t$, respectively. 
Bottom: The temperature dependence of the leading eigenvalues of the BSE equation of the bilayer Hubbard model with $t_{\perp}=1.8t$ and different values of $t_{\perp}^{\prime}$. The remaining model parameters are $U=8t$ and $n=1.15$. 
Both spectral function results were obtained on an $N = 4\times 4\times 2$ cluster, while the eigenvalues of the 
BSE were obtained on an $N = 2\times 2\times 2$ cluster. 
 }
\label{fig:U8}
\end{figure}

Here, we consider results for $n=1.15$, $t_\perp =1.8t$, $\beta=5/t$, but an increased Hubbard interaction $U=8t$. 
The resulting spectral functions are plotted in the left and right panels of Fig. \ref{fig:U8} for $t_\perp^{\prime}=0$ and $0.75t$, respectively. Compared to the case with $U=6t$ (Fig.~\ref{fig:Akwtperprime}, 4\textsuperscript{th} column, first row), we observe that increased $U$ pushes the hole band further below the Fermi level for $t_\perp^{\prime}=0.75t$. The corresponding temperature dependence of the leading BSE eigenvalues, shown in the bottom panel of Fig.~\ref{fig:U8}, clearly shows that T$_c$ is suppressed significantly once the hole-like band becomes incipient.





\bibliography{references}
\end{document}